\documentclass{PoS}

\usepackage{psfrag}
\usepackage{graphicx}

\usepackage{amssymb}
\usepackage{xspace}
\usepackage{epsfig}

\newcommand{\CnuB}{C\(\nu\)B\xspace}
\newcommand{\UHEnu}{UHE\(\nu\)\xspace}
\newcommand{\slache}[1]
{#1\!\!\!\!/}

\title{Interaction of ultra-energetic cosmic neutrinos with a thermal gas of relic neutrinos}

\ShortTitle{Interaction of ultra-energetic cosmic neutrinos with a thermal gas of relic neutrinos}

\author{\speaker{V\'eronique Van Elewyck}\\
        Instituto de Ciencias Nucleares, Universidad Nacional Aut\'onoma de M\'exico\\
	Mexico City, Mexico\\
        E-mail: \email{vero@nucleares.unam.mx}}

\author{J. C. D'Olivo, L. Nellen and S. Sahu\\
        Instituto de Ciencias Nucleares, Universidad Nacional Aut\'onoma de M\'exico\\
	Mexico City, Mexico\\}

\abstract{We use the formalism of finite-temperature field theory to study the interactions of %%@
ultra-high energy (UHE) cosmic neutrinos with the thermal background of relic
neutrinos.
From the imaginary part of the neutrino self-energy, calculated in terms of the Z boson propagator %%@
near the resonance, we derive general expressions for the UHE neutrino transmission probability. %%@
This allows us to take into account the thermal effects introduced by the momentum distribution of %%@
the relic neutrinos.
We compare our results with the approximate expressions existing in the literature and discuss the %%@
influence of thermal effects on the absorption dips in the context of realistic UHE neutrino %%@
fluxes and favoured neutrino mass schemes.}

\FullConference{International Europhysics 
                Conference on High Energy Physics\\
                July 21st - 27th 2005\\
	        Lisboa, Portugal}

\begin{document}

\section{Introduction}

The interaction of cosmic neutrinos at ultra-high energies (\UHEnu) with the cosmological %%@
background of relic (anti)neutrinos (\CnuB) has been proposed as a way of observing the \CnuB, and %%@
a method to perform relic neutrino spectroscopy \cite{UHEnuCnuB}. Provided adequate sensitivity %%@
and energy resolution of the detectors, the observation in the \UHEnu flux of absorption lines %%@
associated with the resonant production of a Z boson ($\nu\bar{\nu} \rightarrow Z \rightarrow f %%@
\bar{f}$) could indeed allow an indirect determination of the absolute neutrino masses. The shape %%@
and depth of these absorption dips may also reflect features of the distribution of \UHEnu sources %%@
and of their emission spectrum \cite{nuspectro}.
Most of the work in the literature describe the \UHEnu -\CnuB interactions assuming that relic %%@
neutrinos are at rest. However, effects of thermal motion in the \CnuB (whose present temperature %%@
is $\approx 1.69 \times 10^{-4}$ eV) become relevant as soon as the momentum of the relic %%@
neutrinos gets comparable to their mass, and even before. To take this effect into account, we %%@
compute the dominant (resonant) contribution to the neutrino damping using the real-time formalism %%@
of finite-temperature field theory (FTFT), and investigate the modifications in the \UHEnu %%@
transmission probability due to thermal effects \cite{ourpaper}. 

\section{Damping of \UHEnu across the \CnuB}
For an \UHEnu with four-momentum $k^\mu = (\mathcal{E}_{_K},\vec{K})$ and mass $m_\nu$ travelling %%@
across the \CnuB, the equation of motion reads $(\slache{k} - m_\nu - \Sigma)\psi = 0$, where the %%@
self-energy $\Sigma$ embodies the effects of the medium. The corresponding dispersion relation is %%@
given by $\mathcal{E}_K = \mathcal{E} _r(K) - i\,\gamma (K)/2$. In our case, $\Sigma$ is %%@
determined from a FTFT one-loop calculation carried out in terms of the (vacuum) Z boson %%@
propagator and the thermal propagator of the relic neurinos.%, having in mind that in the %%@
%real-time formalism of %FTFT, all vertices are doubled and the propagators are
%having in mind that the vertices are doubled respect to the standard theory, and that %the %%@
%propagators of the Z boson (near the resonance) and of the relic neutrino become 
%$2 \times 2$ matrices. 
The last one depends on the functions $f_\nu(P)$ and $f_{\bar{\nu}}(P)$ which describe the %%@
momentum distributions of neutrinos (antineutrinos) in the thermal bath. These functions take the %%@
simple relativistic Fermi-Dirac form, $f_\nu(P) = f_{\bar{\nu}}(P)=1/(e^{P/T_\nu}+1)$, where
$T_{\nu}$ is the temperature of the \CnuB and we have neglected the chemical potential.
 
The damping factor $\gamma$ governs the propagation of the \UHEnu across the background of relic %%@
neutrinos and is directly related to the imaginary part of the self-energy, $\Sigma_{\,i}$ %%@
\cite{dolivo}. In the approximation that the \UHEnu are ultrarelativistic and we can neglect the %%@
background effects on their energy ($\mathcal{E} _r (K)\simeq K$), the damping can be written as %%@
(see \cite{ourpaper} for the detailed calculation)
\begin{equation}
\label{eq:gammaUR} \gamma_{\nu\bar{\nu}}(K) = %%@
-\frac{1}{K}\left.\mathop{\mathrm{Tr}}(\slache{k}\Sigma_i)\right|_{\mathcal{E}_r=K}
= \int_0^\infty \frac{dP}{2\pi^2}
\ P^2 \ f_{\bar{\nu}} (P) \ \sigma_{\nu\bar{\nu}} (P,K).
\label{eq:gamma}
\end{equation}
 For $m_\nu \ll M_Z,K$ and neglecting terms of order $\Gamma_Z^2/M_Z^2$, we have
\begin{eqnarray}
\sigma_{\nu\bar{\nu}}(P,K) &=& \frac{2\sqrt{2}G_\mathrm{F}\Gamma_Z M_Z}{2KE_p}
\left\{ 1 + \frac{M_Z^2}{4KP} \ln\left(\frac{4K^2(E_p + P)^2 -
4M_Z^2K(E_p+P)+M_Z^4}{4K^2(E_p - P)^2 -
4M_Z^2K(E_p-P)+M_Z^4}\right)\right. \nonumber \\
&& \nonumber\\
&& \left. + \frac{M_Z^3}{4KP \Gamma_Z}\left[\arctan\left(\frac{2K(E_p+P)-M_Z^2}{\Gamma
M_Z}\right) -\arctan\left(\frac{2K(E_p-P)-M_Z^2}{\Gamma
M_Z}\right)\right]\right\}, \label{eq:sigma}
\end{eqnarray}
where $E_p = \sqrt{P^2 + m_\nu^2}$ is the energy of the relic neutrino. Taking the limit of %%@
eq.~(\ref{eq:sigma}) for $P \rightarrow 0$, one recovers the approximated cross-section used for %%@
relic neutrinos at rest, with the Z peak at the \UHEnu "bare" resonance energy $K_{res} = %%@
M_Z^2/(2m_\nu)$. However, this approximation breaks down for small $m_\nu$: Fig.~1 (top line) %%@
shows how the resonance peak in the $\nu \bar{\nu} \rightarrow Z$ cross-section broadens and %%@
shifts to lower \UHEnu energies as $P$ increases. The transmission probability for an \UHEnu %%@
emitted at a redshift $z_s$ to be detected on Earth with an energy $K_0$ is obtained by %%@
integrating the damping along the \UHEnu path, taking into account that both the \UHEnu energy and %%@
the \CnuB temperature are redshifted:
\begin{equation}
\label{eq:PTredshift}
P_\mathrm{T}(K_0,z_\mathrm{s})=\exp\left[{-\int_0^{z_\mathrm{s}} \frac{dz}{H(z)(1+z)} %%@
\gamma_{\nu\bar{\nu}}(K_0(1+z))}\right],
\end{equation} 
where $H = H_0\ \sqrt{0.3(1+z)^3 +0.7}$ is the Hubble factor. Fig.~1 (bottom line) shows that for %%@
$m_\nu/T_\nu \lesssim 10^{2}$ the absorption lines are also significantly broadened and shifted to %%@
lower energies, and that the effect increases with the distance travelled by the \UHEnu. This %%@
complicates the extraction of $m_\nu$ and $z_s$ from the start- and endpoint of the absorption %%@
dip, which, in the approximation of relic neutrinos at rest, were respectively located at %%@
$K_0=K_{res}/(1+z_s)$ and $K_0=K_{res}=M_Z^2/(2m_\nu)$. 

\begin{figure}%[ht!]
\label{fig:}
%\begin{center}{}
 \psfrag{20}[c]{\tiny \phantom{n} {$_2$}}
 \psfrag{40}[c]{\tiny \phantom{n} {$_4$}}
 \psfrag{60}[c]{\tiny \phantom{n}{$_6$}}
 \psfrag{80}[c]{\tiny \phantom{n}{$_8$}}
\psfrag{200}[c]{\tiny \phantom{n}{$_2$}}
 \psfrag{400}[c]{\tiny \phantom{n}{$_4$}}
 \psfrag{600}[c]{\tiny \phantom{n}{$_6$}}
 \psfrag{800}[c]{\tiny \phantom{n}{$_8$}}
\psfrag{1000}[c]{\tiny \phantom{n}{}}
\psfrag{2000}[c]{\tiny \phantom{n}{$_2$}}
 \psfrag{4000}[c]{\tiny \phantom{n}{$_4$}}
 \psfrag{6000}[c]{\tiny \phantom{n}{$_6$}}
\psfrag{100}[c]{\tiny \phantom{n}{}}
\psfrag{2}[c]{\tiny \phantom{nn} \raisebox{0.1cm}{$_2$}}
\psfrag{4}[c]{\tiny \phantom{nn} \raisebox{0.1cm}{$_4$}}
\psfrag{6}[c]{\tiny \phantom{nn} \raisebox{0.1cm}{$_6$}}
 \psfrag{1.5}[c]{\tiny \phantom{nnn} \raisebox{0.1cm}{$_{1.5}$}}
 \psfrag{1}[c]{\tiny \phantom{nn} \raisebox{0cm}{$_1$}}
 \psfrag{0.5}[c]{\tiny \phantom{nnn} \raisebox{0cm}{$_{0.5}$}}
 \psfrag{0}[c]{\tiny \raisebox{0.1cm}{$_0$}}
 \psfrag{A}[c]{\tiny \phantom{mmmmmmmmmm}\raisebox{-0.2cm}{$P\,\, [10^{-3}\,\mathrm{eV}]$}}
\psfrag{BBB}[c]{\tiny \phantom{mmmmm}\raisebox{0cm}{$K\,\, [10^{24}\,\mathrm{eV}]$}}
\psfrag{BB}[c]{\tiny \phantom{mmmmm}\raisebox{-0.2cm}{$K\,\, [10^{23}\,\mathrm{eV}]$}}
\psfrag{B}[c]{\tiny \phantom{mmmmm}\raisebox{-0.2cm}{$K\,\, [10^{22}\,\mathrm{eV}]$}}
 \psfrag{27}[c]{\tiny \phantom{n} \raisebox{0cm}{$10^{27}$}}
 \psfrag{26}[c]{\tiny \phantom{n} \raisebox{0cm}{$10^{26}$}}
 \psfrag{25}[c]{\tiny \phantom{n} \raisebox{0cm}{$10^{25}$}}
 \psfrag{24}[c]{\tiny \phantom{n} \raisebox{0cm}{$10^{24}$}}
 \psfrag{23}[c]{\tiny \phantom{n} \raisebox{0cm}{$10^{23}$}}
 \psfrag{22}[c]{\tiny \phantom{n} \raisebox{0cm}{$10^{22}$}}
 \psfrag{21}[c]{\tiny \phantom{n} \raisebox{0cm}{$10^{21}$}}
 \psfrag{1.}[c]{\tiny \phantom{} \raisebox{0.1cm}{$1$}}
 \psfrag{0.8}[c]{\tiny \phantom{} \raisebox{0.1cm}{$0.8$}}
 \psfrag{0.6}[c]{\tiny \phantom{} \raisebox{0.1cm}{$0.6$}}
 \psfrag{0.4}[c]{\tiny \phantom{} \raisebox{0.1cm}{$0.4$}}
 \psfrag{0.2}[c]{\tiny \phantom{} \raisebox{0.1cm}{$0.2$}}
 \psfrag{0.}[c]{\tiny \phantom{} \raisebox{0.1cm}{$0$}}
 \psfrag{X}[c]{\tiny \raisebox{-0.5cm}{$K_0 [\mathrm{eV}]$}}
 \psfrag{Y}[c]{\tiny {$P_\mathrm{T}$}}
\psfig{file=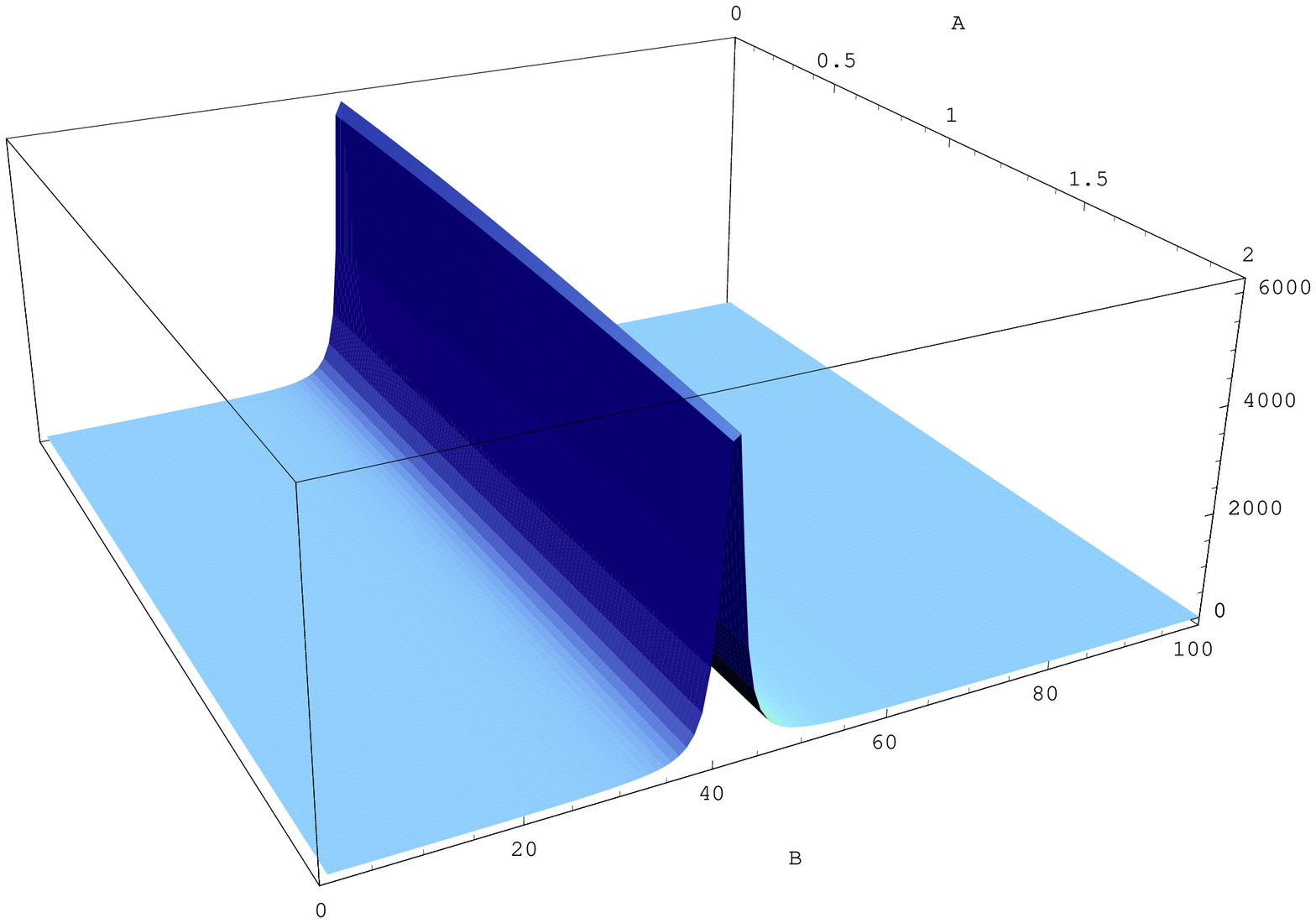,width=.27\textwidth}
 \hfill
 \psfig{file=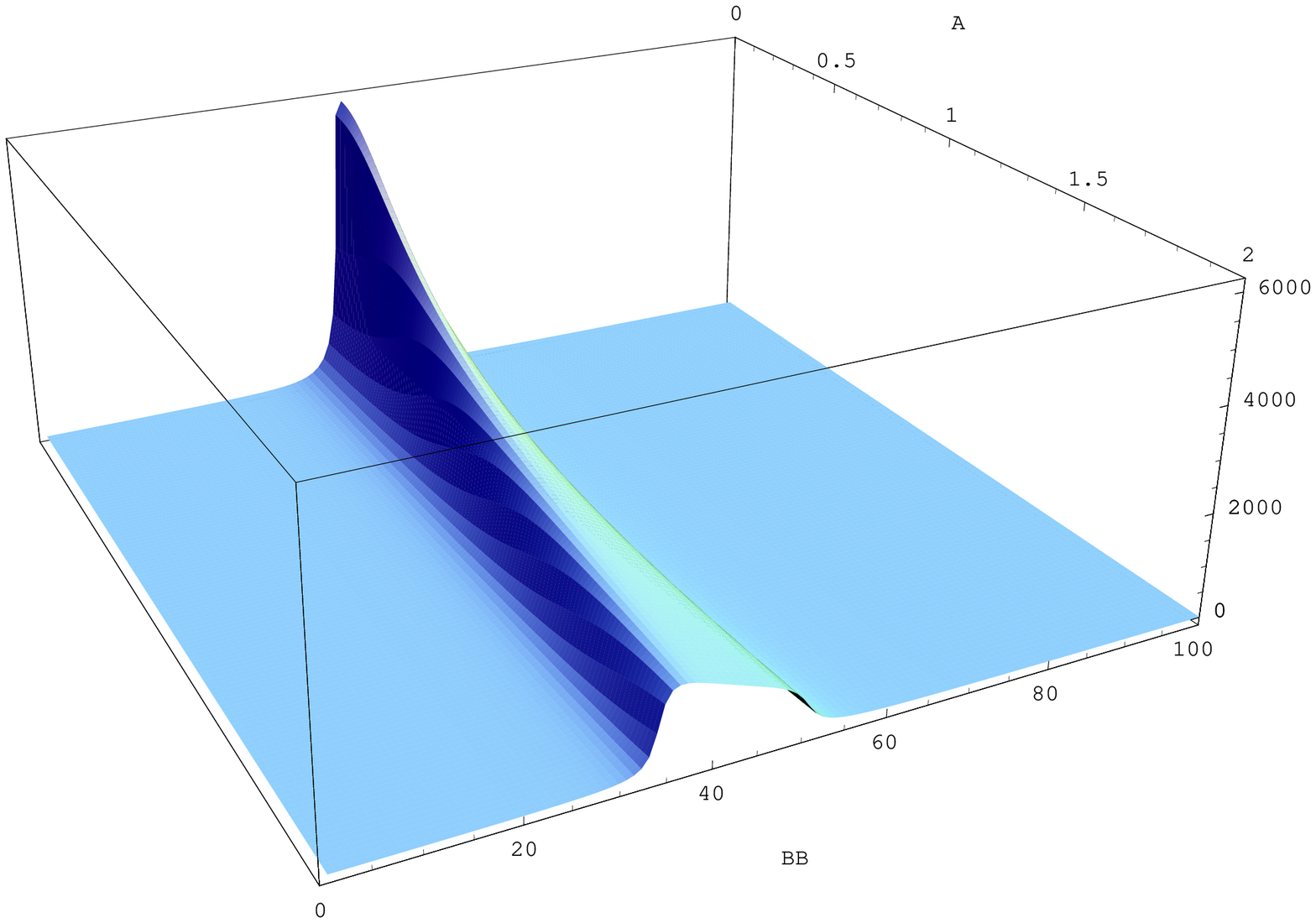,width=.27\textwidth,angle=0}
 \hfill 
 \psfig{file=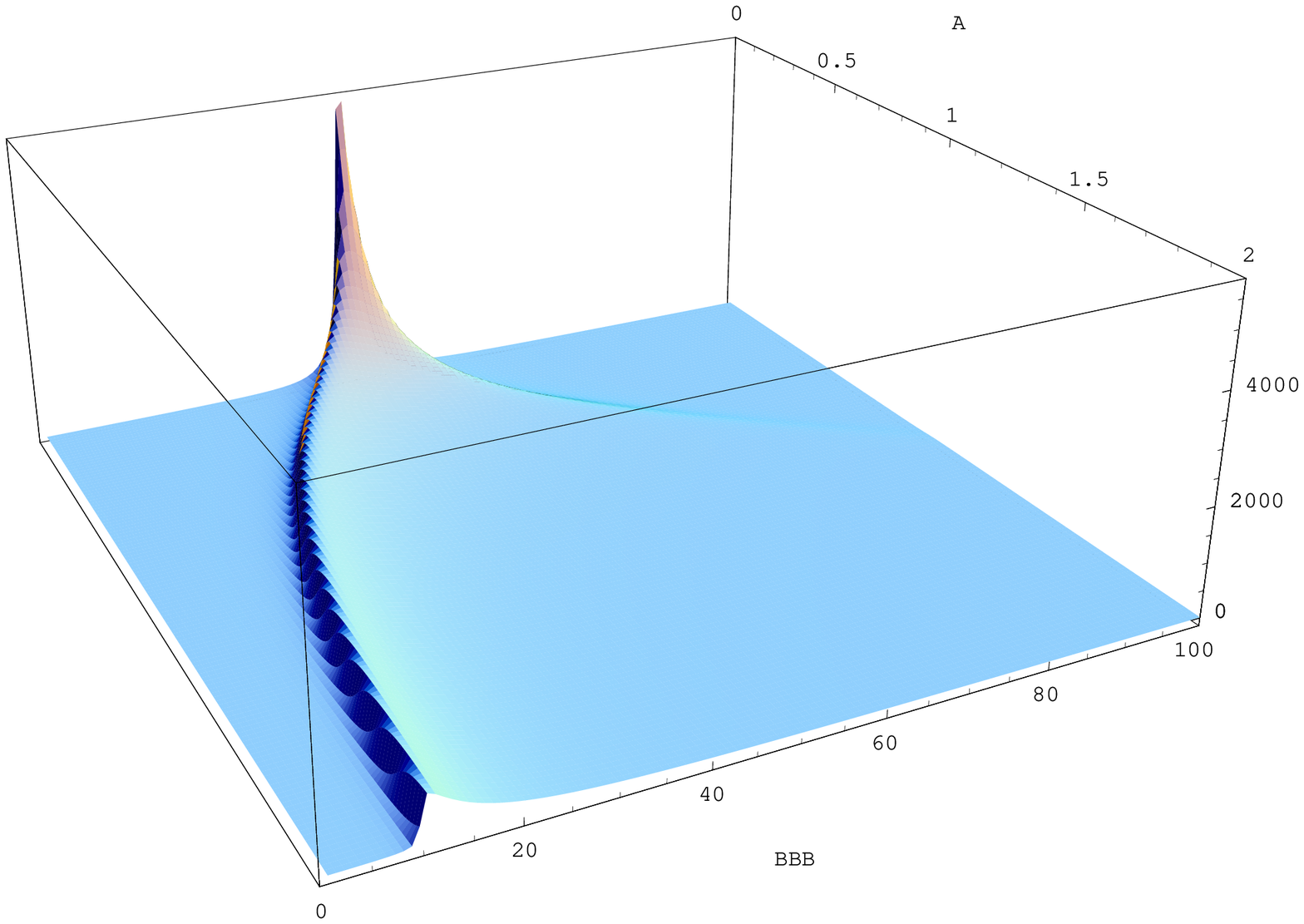,width=.27\textwidth,angle=0} \\[3ex]
 %\hfill 
 \psfig{file=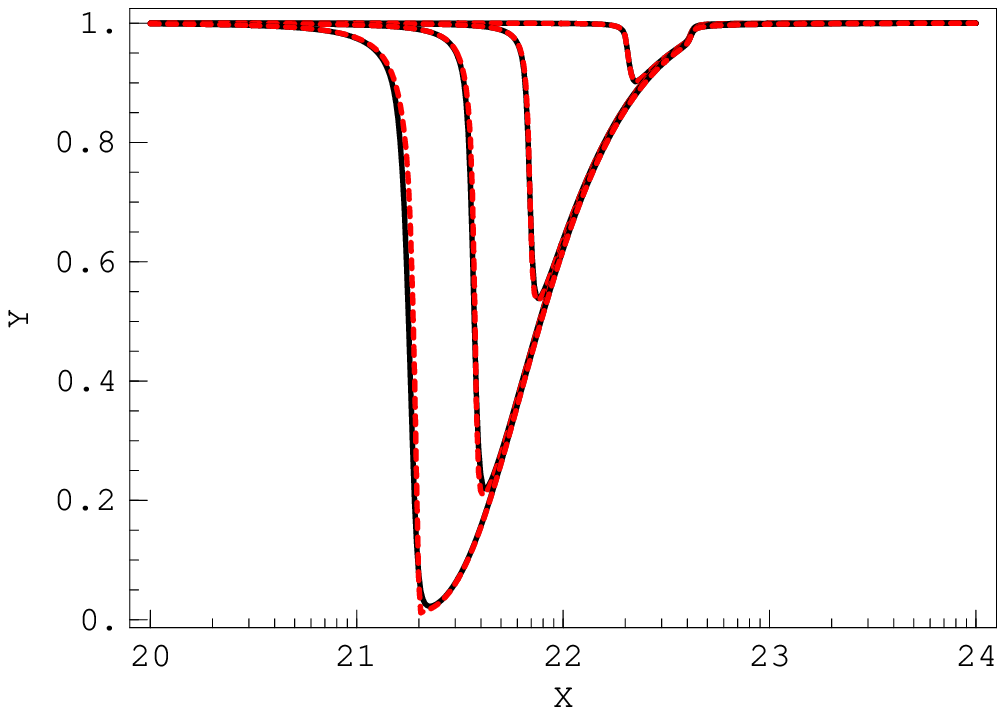,width=0.22\textwidth,angle=0}
 \hfill
 \psfig{file=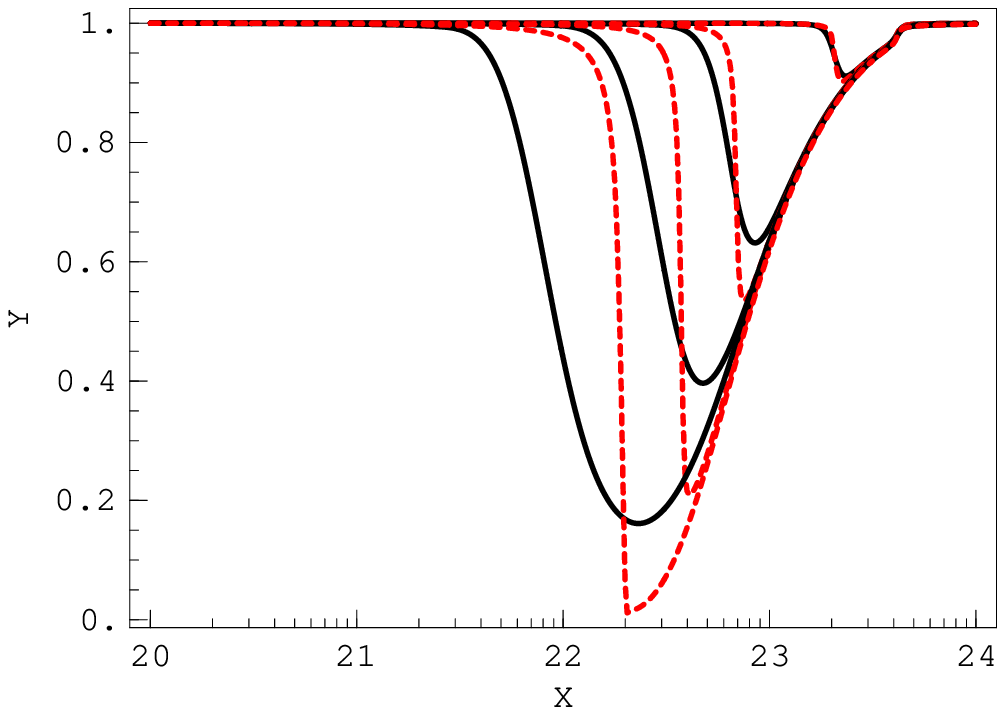,width=0.22\textwidth,angle=0} 
 \hfill
 \psfig{file=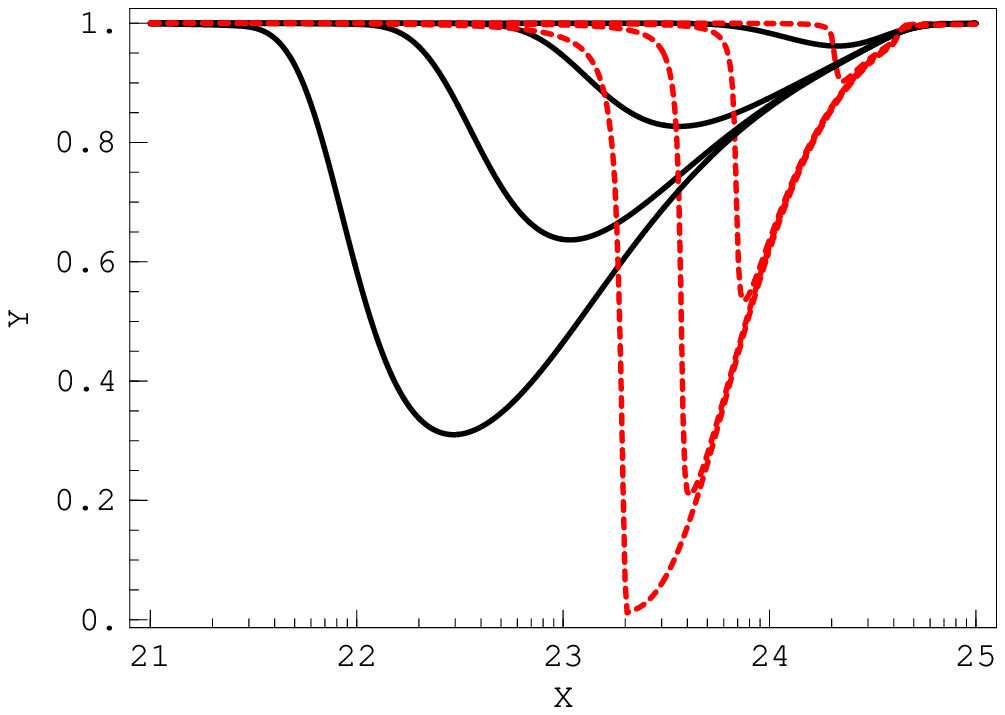,width=0.22\textwidth,angle=0}
% \hfill
 \caption{ \footnotesize{ {\bf Top :} cross-section $\sigma_{\nu\bar{\nu}}(P,K)$, in units of %%@
$10^{-31} cm^2$, as given by eq.~(2.2), as a function of the energy of the incident neutrino, $K$, %%@
and of the relic neutrino momentum, $P$. From left to right, the panels correspond to a neutrino %%@
mass $10^{-1}$, $10^{-2}$, and $10^{-3}$ eV. 
{\bf Bottom :} Transmission probability $P_\mathrm{T}(K_0,z_\mathrm{s})$ as a function of the %%@
\UHEnu energy as detected on Earth, $K_0$, for a source located at redshifts $z_\mathrm{s}=1$, %%@
$5$, $10$, $20$ (from top to bottom in each panel) and for a neutrino mass $m_\nu = 10^{-1}$, %%@
$10^{-2}$, $10^{-3}$ eV (from left to right). The continued, black curves correspond to the full %%@
damping as given by eqs.~(2.1) and (2.2), while the dotted (red) curves are for the approximation %%@
of relic neutrinos at rest.
} 
}
% \end{center}
 \end{figure}

\section{Absorption lines in the \UHEnu flux}
To investigate this effect in a realistic context, we applied our calculation to a flux of \UHEnu 
\begin{equation}
\label{eq:flux}
\mathcal{F}_\nu(K_0) = \frac{1}{4 \pi} \int_0^\infty \frac{dz}{H(z)} \ P_\mathrm{T}(K_0,z) \ %%@
\eta(z)\ J_\nu(K_0),
\end{equation}
assuming a distribution of sources $\eta(z)=\eta_{\,0} \,(1+z)^n\, \theta(z-z_\mathrm{min})\, %%@
\theta(z_\mathrm{max}-z)$ with a common injection spectrum $J_\nu(K)= j_\nu \, K^{-\alpha}\, %%@
\theta(K_\mathrm{max}-K)$. The normalized flux then only depends on the difference of  spectral %%@
indexes, $\alpha-n$, with typically $\alpha-n\approx 2$ for astrophysical (bottom-up) sources and    %%@
$\alpha-n \approx 0$ for top-down processes \cite{nuspectro}. For these two cases, we computed the %%@
normalized, all-flavour \UHEnu spectrum assuming some mass patterns compatible with the currently %%@
favoured three-neutrinos mass schemes \cite{bell}. Fig.~2 shows how thermal broadening affects the %%@
superposition of absorption lines and globally modifies the shape and extension of the dip.   

\begin{figure}%[ht!]
\label{fig:fluxes}
%\begin{center}{}
 \psfrag{27}[c]{\tiny \phantom{n} \raisebox{0cm}{$10^{27}$}}
 \psfrag{26}[c]{\tiny \phantom{n} \raisebox{0cm}{$10^{26}$}}
 \psfrag{25}[c]{\tiny \phantom{n} \raisebox{0cm}{$10^{25}$}}
 \psfrag{24}[c]{\tiny \phantom{n} \raisebox{0cm}{$10^{24}$}}
 \psfrag{23}[c]{\tiny \phantom{n} \raisebox{0cm}{$10^{23}$}}
 \psfrag{22}[c]{\tiny \phantom{n} \raisebox{0cm}{$10^{22}$}}
 \psfrag{21}[c]{\tiny \phantom{n} \raisebox{0cm}{$10^{21}$}}
 \psfrag{20}[c]{\tiny \phantom{n} \raisebox{0cm}{$10^{20}$}}
 \psfrag{1.}[c]{\tiny \phantom{} \raisebox{0.1cm}{$1$}}
 \psfrag{0.9}[c]{\tiny \phantom{} \raisebox{0.1cm}{$0.9$}}
 \psfrag{0.8}[c]{\tiny \phantom{} \raisebox{0.1cm}{$0.8$}}
 \psfrag{0.7}[c]{\tiny \phantom{} \raisebox{0.1cm}{$0.7$}}
 \psfrag{0.6}[c]{\tiny \phantom{} \raisebox{0.1cm}{$0.6$}}
 \psfrag{0.4}[c]{\tiny \phantom{} \raisebox{0.1cm}{$0.4$}}
 \psfrag{0.2}[c]{\tiny \phantom{} \raisebox{0.1cm}{$0.2$}}
 \psfrag{0.}[c]{\tiny \phantom{} \raisebox{0.1cm}{$0$}}
 \psfrag{X}[c]{\tiny \raisebox{-0.5cm}{$K_0 [\mathrm{eV}]$}}
 \psfrag{Y}[c]{\tiny {}}

 \psfig{file=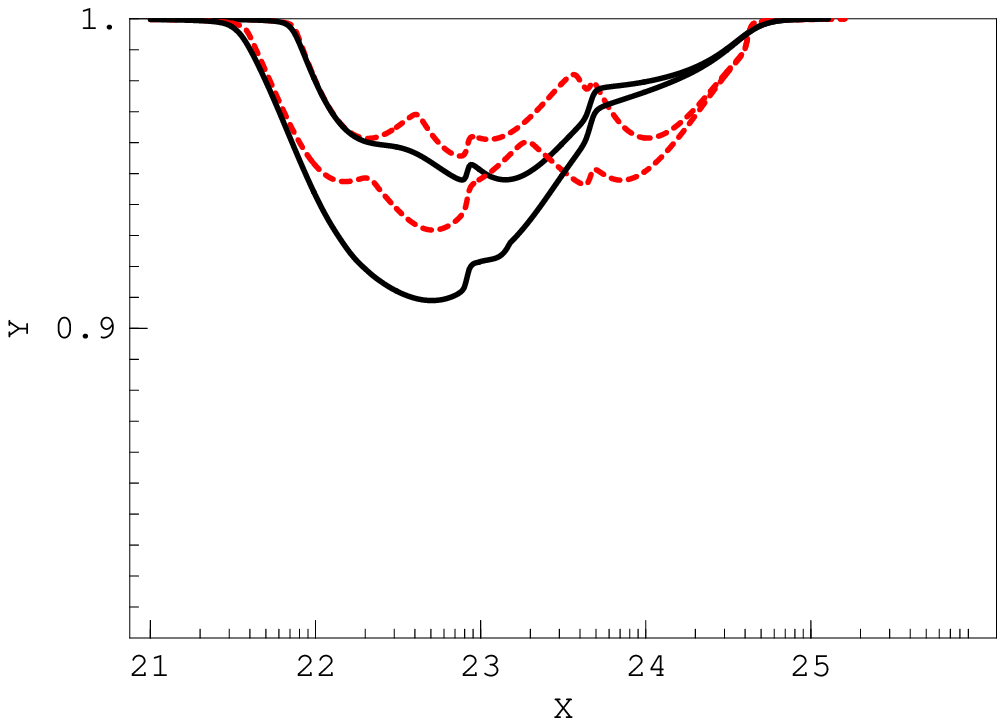,width=0.22\textwidth,angle=0}
 \hfill
 \psfig{file=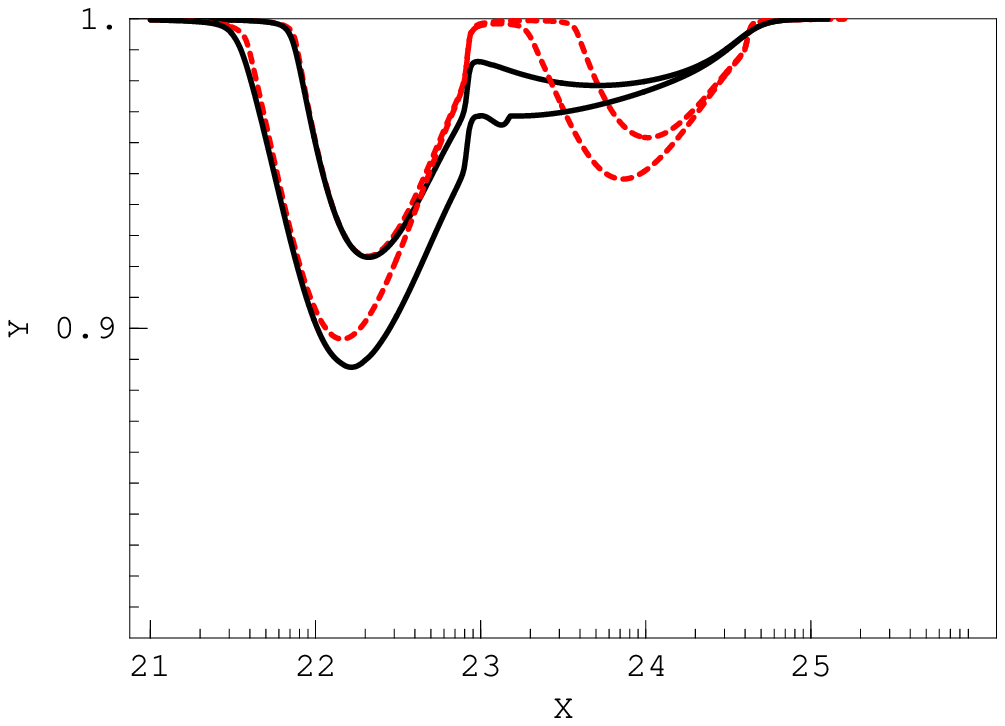,width=0.22\textwidth,angle=0} 
 \hfill
 \psfig{file=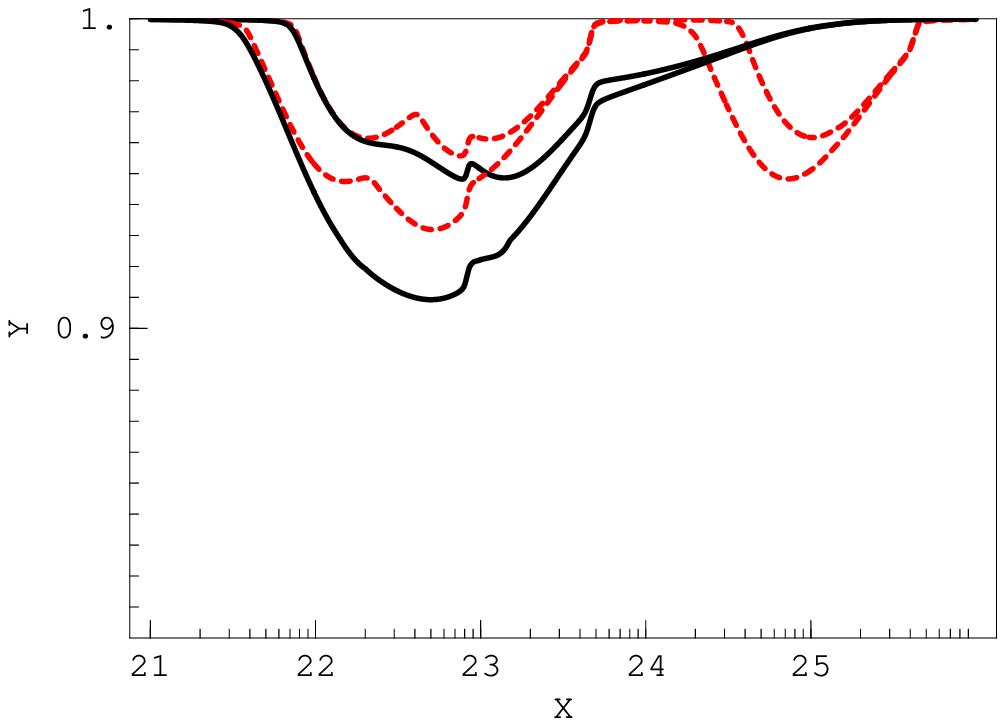,width=0.22\textwidth,angle=0}
 \hfill
 \psfig{file=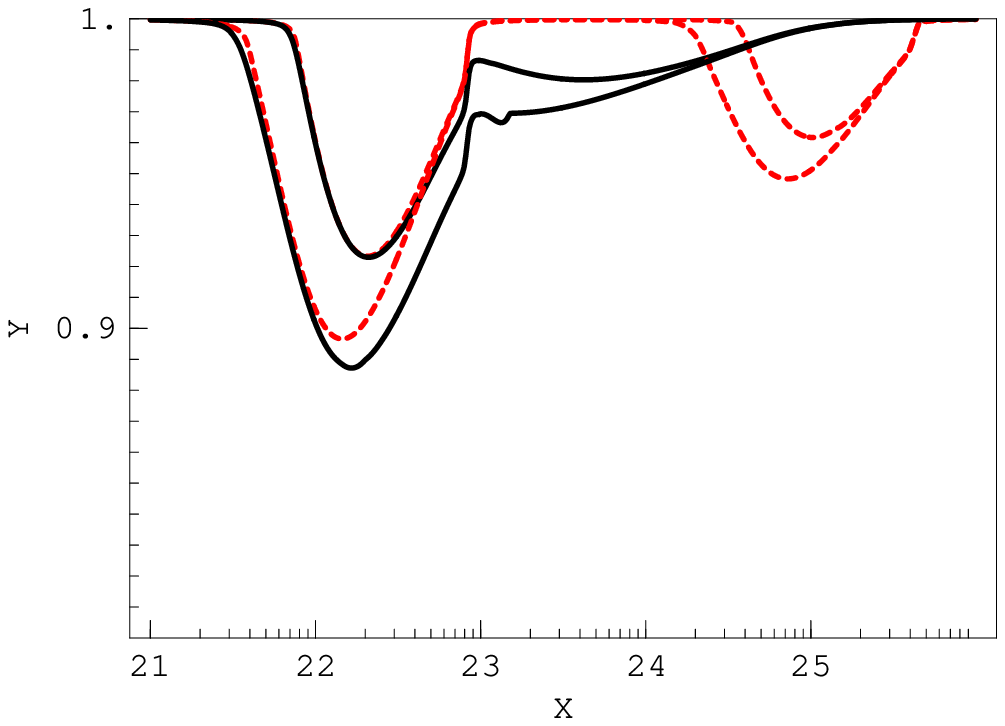,width=0.22\textwidth,angle=0} 
 \hfill \\[3ex]
 \psfig{file=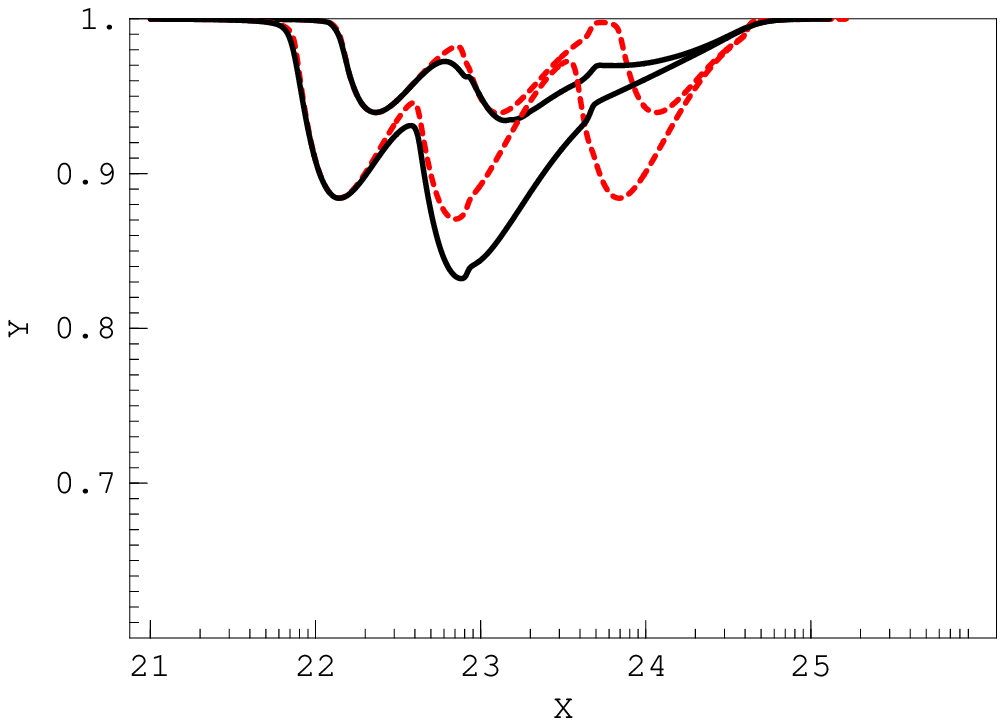,width=0.22\textwidth,angle=0}
 \hfill
 \psfig{file=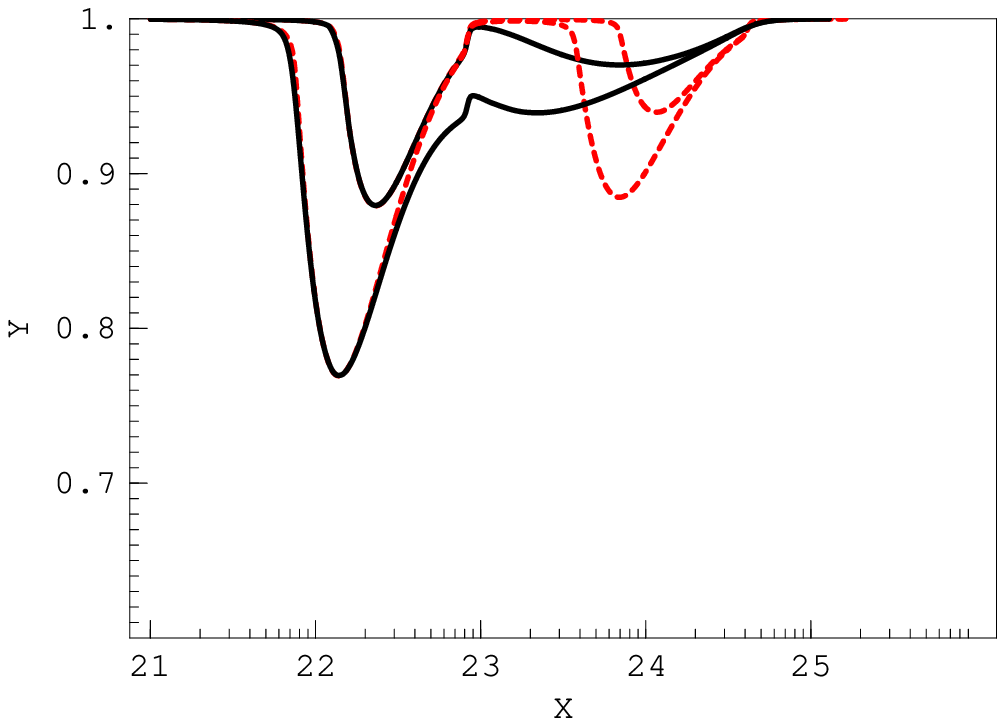,width=0.22\textwidth,angle=0} 
 \hfill
 \psfig{file=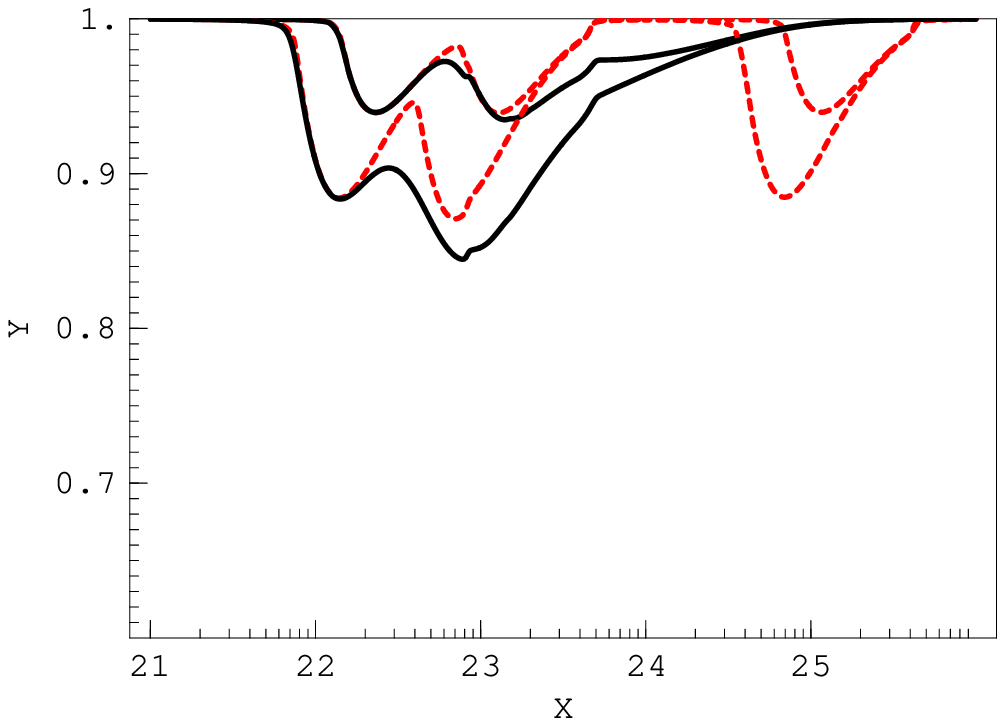,width=0.22\textwidth,angle=0}
 \hfill
 \psfig{file=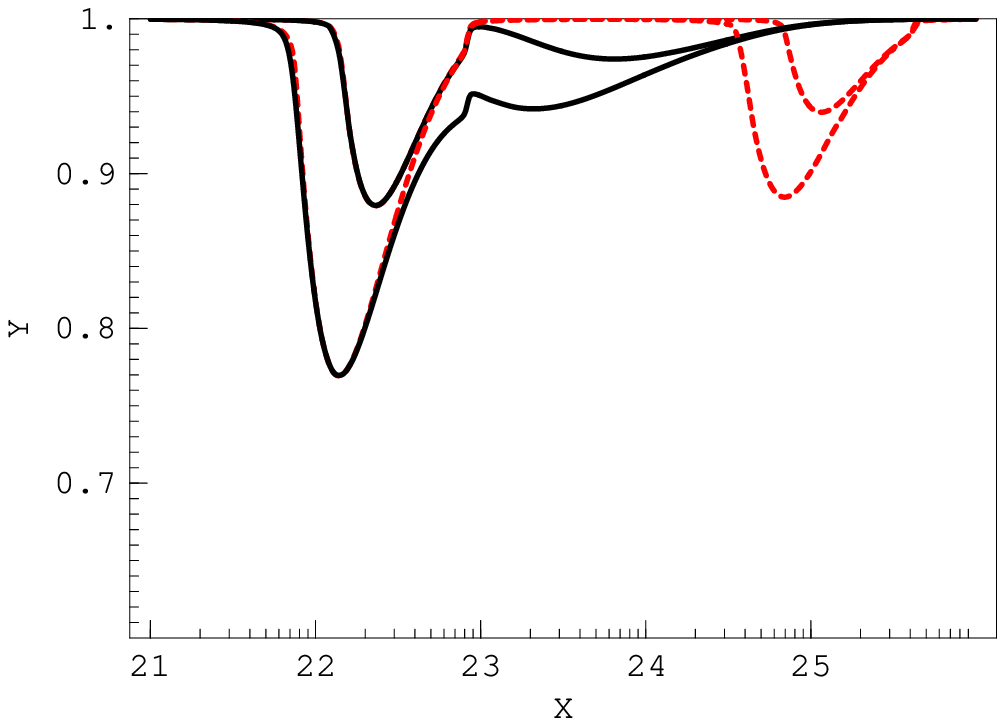,width=0.22\textwidth,angle=0} 
 \caption{ \footnotesize{ \UHEnu flux in presence of damping, $\mathcal{F}_\nu$, normalized to the %%@
corresponding flux in absence of interactions. The top row is for $\alpha-n = 0$ and $z_s = %%@
10,20$, and the bottom row is for $\alpha-n = 2$ and $z=5,10$. From left to right, columns %%@
correspond to the following neutrino mass patterns: $\{10^{-3},\ 9 \times 10^{-3},\ 5 \times %%@
10^{-2}\}$,\ $\{10^{-3},\ 5 \times 10^{-2},\ 5 \times 10^{-2}\}$,\ $\{10^{-4},\ 9 \times 10^{-3},\  %%@
5 \times 10^{-2}\}$,\ $\{10^{-4},\ 5 \times 10^{-2},\ 5 \times 10^{-2}\}$ (all in eV). Colour code %%@
is as in fig. 1.}} 
% \end{center}
 \end{figure}

\section{Conclusions}
From the exploration of the parameter space currently allowed by astrophysical and cosmological %%@
constraints, we see that thermal effects do affect the transmission properties of \UHEnu across %%@
the \CnuB even in the regime of non-relativistic relic neutrinos. For most neutrino mass patterns, %%@
the extraction of the neutrino masses from the endpoint of the absorption lines is complicated by %%@
the broadening and merging of the dips, especially in normal hierarchical schemes (columns 1 and 3 %%@
in fig.~2) and for top-down-like injection spectra. Some information on the source distribution %%@
could still be provided by the onset energy and slope of the dip.

\section*{Acknowledgements}
We would like to acknowledge support by CONACyT under grants 34868-E
and 46999-F and by DGAPA-UNAM under grants PAPIIT IN116503, IN119405,
and IN112105.


\begin{thebibliography}{99}
%\bibitem{langacker83}P.~Langacker, J.~P.~Leveille, and J.~Sheiman,
%``On The Detection Of Cosmological Neutrinos By Coherent Scattering,''
%Phys.\ Rev.\ D {\bf 27}, 1228 (1983).
\bibitem{UHEnuCnuB} 
T.~J.~Weiler,
%``Resonant Absorption Of Cosmic Ray Neutrinos By The Relic Neutrino
%Background,''
Phys.\ Rev.\ Lett.\  {\bf 49}, 234 (1982) and 
%``Big Bang Cosmology, Relic Neutrinos, And Absorption Of Neutrino Cosmic
%Rays,''
Astrophys.\ J.\  {\bf 285}, 495 (1984); E.~Roulet,
%``Ultrahigh-energy neutrino absorption by neutrino dark matter,''
Phys.\ Rev.\ D {\bf 47}, 5247 (1993); 
H.~Pas and T.~J.~Weiler,
  %``Absolute neutrino mass determination,''
  Phys.\ Rev.\ D {\bf 63} (2001) 113015; D.~Fargion, P.~G.~De Sanctis Lucentini, M.~Grossi, M.~De %%@
Santis, and B.~Mele,
  %``Ultra high energy cosmic ray and UHE nu nu(r) Z showering in dark  halos,''
  Mem.\ Soc.\ Ast.\ It.\  {\bf 73} (2002) 848.  

\bibitem{nuspectro}
B.~Eberle, A.~Ringwald, L.~Song, and T.~J.~Weiler,
%``Relic neutrino absorption spectroscopy,''
Phys.\ Rev.\ D {\bf 70}, 023007 (2004); G.~Barenboim, O.~Mena Requejo, and C.~Quigg,
%``Diagnostic potential of cosmic-neutrino absorption spectroscopy,''
Phys.\ Rev.\ D {\bf 71} (2005) 083002.  
\bibitem{bell} J.~F.~Beacom and N.~F.~Bell,
  %``Do solar neutrinos decay?,''
  Phys.\ Rev.\ D {\bf 65} (2002) 113009.
\bibitem{dolivo} J.~C.~D'Olivo and J.~F.~Nieves,
  %``Damping rate of a fermion in a medium,''
  Phys.\ Rev.\ D {\bf 52} (1995) 2987.
\bibitem{ourpaper} J.~C.~D'Olivo, L.~Nellen, S.~Sahu and V.~Van~Elewyck, arXiv:astro-ph/0507333, %%@
to appear in Astropart. Phys.
  
\end{thebibliography}
\end{document}